\documentclass{jpsj3}
\usepackage{txfonts}
\usepackage{color}
\usepackage{braket}
\usepackage{bm}
\usepackage{xspace}
\usepackage[dvipdfmx]{graphicx}

\newcommand{\be}{\begin{equation}}
\newcommand{\ee}{\end{equation}}
\newcommand{\bea}{\begin{eqnarray}}
\newcommand{\eea}{\end{eqnarray}}
\newcommand{\beaa}{\begin{eqnarray*}}
\newcommand{\eeaa}{\end{eqnarray*}}

\title{Anomalous Phase-Coherence Scaling in a Quantum-Critical Dirac Semimetal}

\author{Sana Nakamichi$^1$, Ryotaro Kobara$^1$, Yoshinari Unozawa$^1$, Yoshitaka Kawasugi$^1$, Sakura Hiramoto$^2$, Koki Funatsu$^2$, Toshio Naito$^2$, Masafumi Tamura$^3$, Reizo Kato$^4$, Yutaka Nishio$^1$, and Naoya Tajima$^1$\thanks{naoya.tajima@sci.toho-u.ac.jp}}

\inst{
$^1$Department of Physics, Toho University, Funabashi, Chiba 274-8510, Japan\\ 
$^2$Graduate School of Science and Engineering, Ehime University, Matsuyama 790-8577, Japan \\
$^3$Department of Physics, Faculty of Science and Technology, Tokyo University of Science, Noda, Chiba 278-8510, Japan\\
$^4$RIKEN, Hirosawa 2-1, Wako-shi, Saitama 351-0198, Japan
}

\abst{
We have investigated the weak antilocalization (WAL) in the pressurized Dirac semimetal $\alpha$-(BEDT-TTF)$_2$I$_3$ across a correlation-driven quantum phase transition to a charge-ordered insulating state and evaluated the phase coherence length $L_{\phi}$ and its temperature scaling under various pressures from the low-temperature magnetoconductivity. In the high-pressure regime, the system exhibits the conventional two-dimensional dephasing behavior ($L_{\phi} \propto T^{-p}$ with $p \approx 1/2$), characteristic of electron–electron scattering in diffusive conductors. As the pressure approaches the critical pressure ($P_c \sim 1.2$ GPa), the temperature exponent is suppressed to $p \sim 0.3$, while $L_{\phi}$ remains large ($700\text{–}800$ nm at 0.5 K). 
This anomalous scaling suggests nontrivial inelastic scattering associated with Dirac electrons near the quantum critical point. The persistence of WAL across the transition supports a gapless or nearly gapless quantum phase transition. 
}

\allowdisplaybreaks

\usepackage{color}

\usepackage{ulem}

\begin{document}
\maketitle

Since the discovery of graphene \cite{rf:1, rf:2}, materials hosting massless Dirac and Weyl fermions have attracted considerable attention as platforms for exploring relativistic quasiparticles in solids. These systems are characterized by linear band dispersions, where the conduction and valence bands intersect at discrete Dirac points in momentum space. The low-energy excitations obey a Dirac-like equation and possess a nontrivial $\pi$ Berry phase, leading to unconventional transport phenomena.

Electron–electron interactions can substantially modify the low-energy physics of Dirac fermion systems. 
Field-theoretical studies based on the Gross–Neveu and Gross–Neveu–Yukawa models predict correlation-driven quantum phase transitions, typically associated with symmetry breaking and mass generation. At the critical point, the system exhibits gapless, strongly interacting excitations \cite{rf:3, rf:4}.
Near such quantum critical points (QCPs), Dirac fermions acquire anomalous scaling dimensions and exhibit non-Fermi-liquid scattering rates \cite{rf:5, rf:6, rf:7}. While spectroscopic signatures of interacting Dirac states have been extensively discussed, experimental transport probes of quantum-critical Dirac regimes, particularly those involving quantum coherence, remain limited.

The organic conductor $\alpha$-(BEDT-TTF)$_2$I$_3$ provides a testing ground for the investigation of this problem. Under pressure, this material realizes a massless Dirac electronic structure with the Fermi level located at the Dirac point \cite{rf:8, rf:9, rf:10, rf:11, rf:12}. Previous studies have demonstrated that at high pressure, $\alpha$-(BEDT-TTF)$_2$I$_3$ exhibits a three-dimensional (3D) Dirac semimetal (DS) state with unconventional magnetotransport signatures \cite{rf:13, rf:14, rf:15}.  Moreover, it is situated in proximity to a charge-ordered insulating phase, indicating the possible presence of strong electron correlations \cite{rf:16, rf:17}. Notably, we have demonstrated that this system undergoes a quantum phase transition to a charge-ordered phase at $P_c \sim 1.2$ GPa without opening a mass gap \cite{rf:18}. These results suggest the emergence of a correlation-driven quantum critical Dirac regime near $P_c$.

In such a regime, quantum interference effects are expected to be strongly influenced by Dirac electrons near the quantum critical point. The weak antilocalization (WAL), arising from coherent backscattering in the presence of spin–orbit coupling, provides a sensitive probe of phase coherence and dephasing mechanisms. Although WAL-like magnetoresistance has been reported in $\alpha$-(BEDT-TTF)$_2$I$_3$ under high pressure \cite{rf:15}, how the quantum-critical regime of the massless Dirac state affects the temperature scaling of the phase coherence length $L_{\phi}$ remains an open question.

In conventional diffusive metals, the temperature dependence of the coherence length $L_{\phi}$ due to electron-electron interactions follows a power-law $L_{\phi} \propto T^{-p}$. For 2D, $p=1/2$, and for 3D, $p=3/4$ \cite{rf:19, rf:20}. In contrast, theory predicts that massless Dirac fermions can renormalize the dephasing dynamics and suppress the exponent $p$ \cite{rf:4, rf:5}. 

In this study, we have investigated low-temperature magnetotransport measurements under pressure and quantitatively analyzed WAL magnetoconductivity. We demonstrate that while $L_{\phi}$ remains remarkably large across the transition, its temperature exponent $p$ exhibits a pronounced suppression near $P_c$, providing transport evidence for a quantum-critical Dirac regime governed by strong correlations.

In the experiments, a sample with electrical leads attached was sealed in a Teflon capsule filled with a pressure medium (Idemitsu DN-oil 7373). The capsule was set in a clamp-type pressure cell with a double-layer structure made of CuBe and NiCrAl hard alloys.

Electrical resistivity $\rho_{xx}$ and Hall resistivity $\rho_{xy}$ were measured over a pressure range of 1.05 GPa to 1.7 GPa and temperatures below 4.2 K using a conventional six-probe DC technique. An electrical current of 0.1 to 10 $\mu$A was applied in the $ab$ plane, and a magnetic field ranging from -0.2 to 0.2 T was applied perpendicular to the $ab$ plane.

\begin{figure}[htbp]
  \includegraphics[width=0.7 \linewidth]{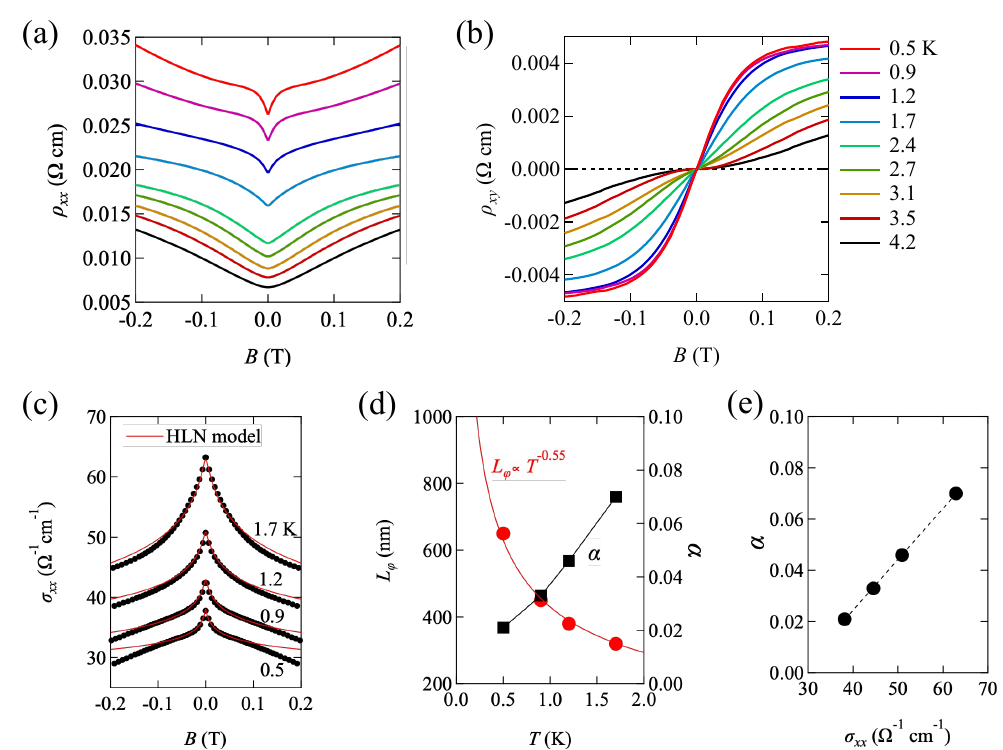}
  \caption{
    \label{fig:1}
(color online) Magnetic field dependence of (a) $\rho_{xx}$ and (b) $\rho_{xy}$ under a pressure of 1.7 GPa at temperature below 4.2 K. (c) $\sigma_{xx}$ estimated from (a) $\rho_{xx}$ and (b) $\rho_{xy}$. The data are fitted using the HLN model. (d) Temperature dependence of the phase coherence length $L_{\phi}$ and the prefactor $\alpha$, which serve as fitting parameters in the HLN model. (e) The prefactor $\alpha$ against the conductivity $\sigma_{xx}$ in the zero field.
  }
\end{figure}

Figures 1(a) and 1(b) show the magnetic-field dependence of the longitudinal resistivity $\rho_{xx}$ and the Hall resistivity $\rho_{xy}$ under a pressure of 1.7 GPa at temperatures below 4.2 K. Notably, $\rho_{xx}$ exhibits a cusp-like feature in the vicinity of zero magnetic field at temperatures below 1.7 K. Such a characteristic low-field anomaly is a hallmark of WAL. In contrast, no corresponding anomaly is observed in $\rho_{xy}$ within the same field and temperature range. This is consistent with the fact that weak localization corrections predominantly affect the longitudinal conductivity rather than the Hall response \cite{rf:19}.

This WAL-like behavior of the magnetoresistance is in good agreement with our previous results \cite{rf:21}. Furthermore, a similar cusp-like feature in $\rho_{xx}$ is observed at temperatures below 1.6 K even when the magnetic field is applied parallel to the $ab$ plane \cite{rf:21}. This anisotropy-independent behavior suggests the emergence of 3D WAL in this system below 1.7 K.

Furthermore, the Dirac state under high pressure exhibits coherent interlayer transport and 3D band dispersion. However, the extremely large conductivity anisotropy ($>10^3$) suggests that quantum interference may be effectively confined within conducting layers. We therefore analyze the magnetoconductivity $\sigma_{xx}= \rho_{xx} / (\rho_{xx}^2 + \rho_{xy}^2)$ using the quasi-2D Hikami-Larkin-Nagaoka (HLN) model \cite{rf:22}.

In Fig. 1(c), $\sigma_{xx}(B)$ is analyzed in terms of the HLN formula normalized by the interlayer spacing $c = 1.7$ nm \cite{rf:23}, expressed as,
\begin{equation}
\sigma_{xx}(B)=\sigma_{xx}(0)
-\alpha\frac{e^2}{2\pi^2\hbar c}\left[\psi\left(\frac{1}{2}+\frac{B_\phi}{B}\right)-\ln\left(\frac{B_\phi}{B}\right)\right],
\label{eqn:1}
\end{equation}
where $\psi(x)$ denotes the digamma function, and the prefactor $\alpha (<1)$ characterizes the effective number of coherent transport channels contributing to quantum interference. $B_{\phi}=\hbar/4eL_{\phi}^{2}$ is the dephasing field with $L_{\phi}=\sqrt{D\tau_{\phi}}$ being the dephasing length, $D$ is the diffusion constant, $\tau_{\phi}$ is the dephasing time.

Equation (1), originally derived for diffusive transport in conventional 2D systems, has also been widely applied to systems with nontrivial band topology, where quantum interference effects are modified by Berry phase and spin-orbit coupling (SOC). 
This simplified HLN formula, without an explicit spin-orbit term, can be justified to the strong SOC limit.
This condition is expected to hold in this system at temperatures below 1.7 K, as first-principles calculations by S. M. Winter et al. indicate that the spin-orbit interaction energy is approximately $1\text{–}2$ meV \cite{rf:24}.
On the other hand, T. Osada suggests that the interlayer SOC, arising from the I$_3^{-}$ anion potential, plays a significant role in the 3D Dirac semimetallic state and in chiral transport phenomena \cite{rf:25, rf:26}.
As shown below, the low-field magnetoconductivity is well described by the HLN formula, enabling reliable extraction of the phase coherence length.

In the present system, as in graphene, since two Dirac cones exist as valleys in the first Brillouin zone, the short-ranged intervalley scattering cannot be ignored. In this case, the magnetoconductivity must be described by an equation that extends the HLN theory derived by McCann et al. \cite{rf:27}.
However, it is difficult to distinguish those contributions uniquely to the observed magnetoconductivity. 
From an experimental viewpoint, we adopt here the HLN Eq. (1) as an effective phenomenological description of the low-field behavior, where the coefficient $\alpha$ captures the net contribution of the quantum interference involving the SOC and intervalley scattering effects.

The analysis is performed in a low-field region where a characteristic cusp structure is dominant. While $\alpha$ reflects the coexistence of multiple scattering processes, the phase coherence length $L_\phi$ is determined primarily by the field scale of the cusp.

Furthermore, in the presence of additional scattering channels, the characteristic field scale may be influenced by multiple contributions, which can lead to an effective reduction of the extracted $L_\phi$ within the HLN framework. Thus, the obtained $L_\phi$ likely provides a lower limit for the true phase coherence length. This robustness reinforces the importance of spin–orbit interaction in the present system, as will be discussed in the final section. Notably, WAL behavior is observed even in the charge-ordered phase, where the electronic structure is significantly modified. This indicates that spin–orbit interaction remains operative across different electronic phases and plays a fundamental role in governing quantum interference in this system.


Now, we analyze the magnetoconductivity using the HLN formula.
The good agreement of $\sigma_{xx}(B)$ with the HLN formula, with a small prefactor $\alpha < 1$ as shown in Fig. 1(d), suggests that the quantum interference responsible for WAL is effectively confined within the conducting layers. The temperature dependence of the phase coherence length $L_{\phi}$ is also shown in Fig. 1(d). The phase coherence length reaches approximately 700 nm at 0.5 K, indicating well-developed quantum interference. The observed relationship, $L_{\phi} \propto T^{-0.55}$, is consistent with the $T^{-0.5}$ dependence expected for the Nyquist electron–electron scattering in 2D diffusive systems, indicating that electron–electron interactions dominate the dephasing. We note that the WAL cusp becomes less pronounced above $\sim$2 K even at 1.7 GPa. This behavior can be understood as a consequence of the reduction in the phase-coherence length $L_{\phi}$ with increasing temperature. Although the spin–orbit scattering length $L_{SO}$ is expected to remain shorter than $L_{\phi}$ at low temperatures, the condition $L_{\phi} \gg L_{SO}$ required for clear WAL behavior may no longer be well satisfied at higher temperatures.

The observed scaling relationship $\alpha \propto \sigma_{xx}(0)$ in Fig. 1(e) suggests that $\alpha$ represents the effective weight of coherent transport channels involved in quantum interference, rather than being an independent fitting parameter. This allows for the self-consistent evaluation of $L_{\phi}$.

Now, let us investigate how the temperature dependence of $L_{\phi}$, $ L_{\phi}\propto T^{-p}$, changes as the system undergoes a quantum phase transition to a charge-ordered insulating phase at approximately 1.2 GPa by increasing electron correlation (releasing pressure) \cite{rf:4, rf:5}.

\begin{figure}[htbp]
  \includegraphics[width=1 \linewidth]{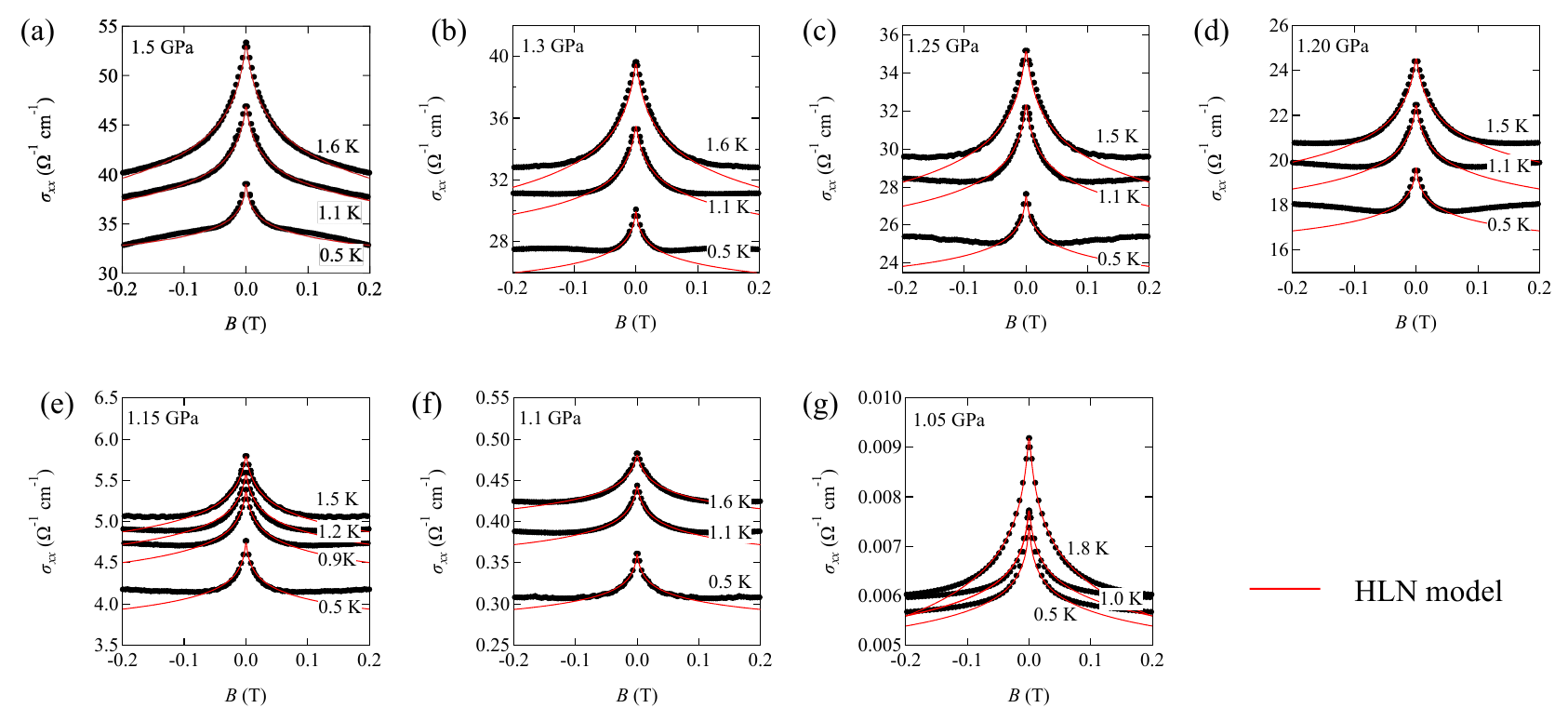}
  \caption{
    \label{fig:2}
  (color online) Magnetic field dependence of $\sigma_{xx}$ at pressures 1.5 GPa (a), 1.3 GPa (b), 1.25 GPa (c), 1.2 GPa (d), 1.15 GPa (e), 1.1 GPa (f), and 1.05 GPa (g). The solid lines represent the fit using the HLN formula in Eq. (1), and the fitting parameters $L_{\phi}$ and $\alpha$ are shown in Fig. 3. 
  }
\end{figure}

\begin{figure}[htbp]
  \includegraphics[width=0.7 \linewidth]{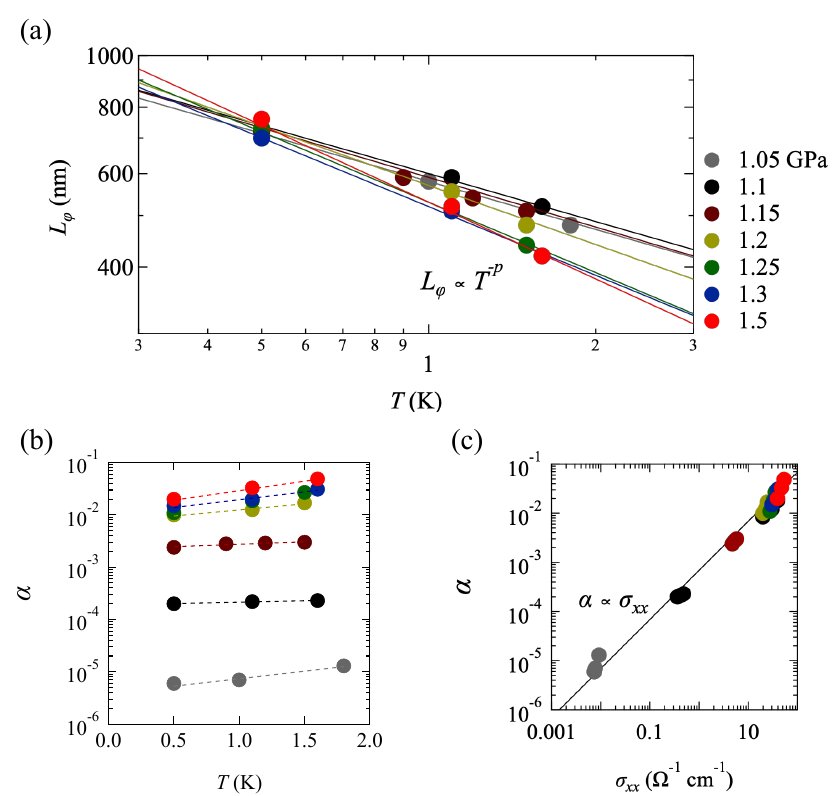}
  \caption{
    \label{fig:3}
  (color online) The temperature dependence of $L_{\phi}$ (a) and $\alpha$ (b) under different pressures. (c) $\alpha$ against $\sigma_{xx}(0)$ under several pressures and temperatures. 
  }
\end{figure}

Figure 2 shows the magnetic field dependence of $\sigma_{xx}$ under different pressures. Using Eq. (1), $L_{\phi}$ and $\alpha$ are evaluated by fitting $\sigma_{xx}$ in the low magnetic field region. The temperature dependence of $L_{\phi}$ is shown in Fig. 3 (a) and that of $\alpha$ in Fig. 3(b). $L_{\phi}$ at each pressure follows a power-law $L_{\phi} \propto T^{-p}$. 
The value of $\alpha$ decreases by approximately four orders of magnitude from 1.5 GPa to 1.05 GPa through the phase transition, in spite of the weak temperature dependence. The relationship of $\alpha \propto \sigma_{xx}(0)$ as shown in Fig. 3(c) demonstrates the validity of evaluating $L_{\phi}$.

\begin{figure}[htbp]
  \includegraphics[width=0.7 \linewidth]{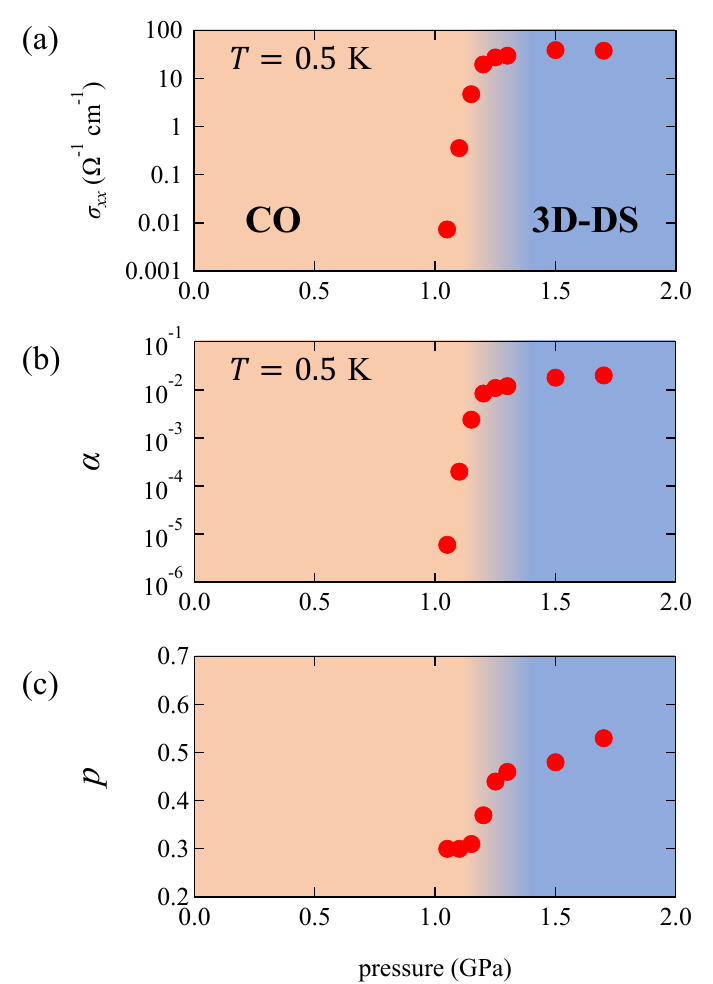}
  \caption{
    \label{fig:4}
  (color online) The pressure dependence of (a) $\sigma_{xx}(0)$ and (b) $\alpha$ at 0.5 K, and (c) the temperature exponent $p$ in the equation $L_{\phi} \propto T^{-p}$.
  }
\end{figure}

The pressure dependence of $\sigma_{xx}(0)$ and $\alpha$ at 0.5 K, and the temperature exponent $p$ in $L_{\phi} \propto T^{-p}$ shown in Fig. 4, reveal a nontrivial critical behavior of the massless Dirac state near the gapless quantum phase transition at $P_c \sim 1.2$ GPa.

In the high-pressure Dirac semimetal regime, the exponent $p$ shown in Fig. 4(c) is close to 0.5–0.55, consistent with dephasing dominated by electron–electron interactions in a diffusive 2D system. The phase coherence length reaches $L_{\phi} \sim 700\text{–}800$ nm at 0.5 K, as shown in Fig. 3(a), indicating well-developed quantum interference of Dirac carriers.

As pressure decreases to $P_c$, a pronounced suppression of $p$ is observed. It reaches a minimum of approximately 0.3 at 1.05–1.2 GPa, as shown in Fig.4 (c). Such a reduction is difficult to understand with conventional diffusive Fermi-liquid dephasing mechanisms and suggests an additional inelastic scattering channel associated with enhanced low-energy fluctuations. The anomalous scaling of $L_{\phi}$ can be interpreted as a manifestation of Dirac criticality near the quantum phase transition, if the transition occurs without opening a mass gap. Notably, $L_{\phi}$ remains as large as $700\text{–}800$ nm at 0.5 K and shows little pressure dependence despite the significant change in the temperature exponent $p$. This indicates that the quantum coherence of Dirac carriers is preserved, whereas the dynamical scaling of dephasing is strongly modified.

The expected transition between the Dirac semimetal and the charge-ordered phase can be described in terms of the Gross–Neveu–Yukawa framework, in which gapless Dirac fermions couple to a critical order-parameter field. At such a quantum critical point, both fermionic and bosonic excitations remain gapless and exhibit relativistic scaling. In the related quantum-critical V-shaped regime extending to finite temperatures, temperature becomes the dominant energy scale, leading to anomalous scaling of the fermionic self-energy and dephasing dynamics \cite{rf:28, rf:29}. The observed suppression of the exponent $p$ near $P_c$ is consistent with this scenario. It implies that the system enters a quantum-critical Dirac regime where interactions change how phase coherence decays with temperature, without strongly suppressing the coherence itself.

The persistence of WAL even in the charge-ordered phase further supports a gapless or nearly gapless character of the transition \cite{rf:18}. The survival of WAL across $P_c$ implies that Dirac-like excitations remain operative on both phases of the transition, consistent with a continuous quantum phase transition governed by quantum-critical fluctuations associated with the charge-order instability rather than a conventional gap-opening mechanism.

Collective consideration of the pressure dependence of $\sigma_{xx}$ and $\alpha$ at 0.5 K, and $p$ shown in Fig. 4, supports the following scenario.
At high pressure, the system is in a stable Dirac semimetal characterized by conventional electron-electron dephasing. 
Near $P_c$, a quantum-critical V-shaped regime emerges, where anomalous scaling of the exponent $p$ ($p \sim 0.3$) indicates enhanced quantum-critical fluctuations associated with the charge-ordered phase transition. The persistence of Dirac interference effects even in the charge-ordered phase suggests that the massless Dirac character is largely preserved across the transition, consistent with a nearly gapless quantum phase transition.
Interestingly, anomalous Hall transport in the same system exhibits an unconventional scaling relation, $\sigma_{xy}^{AHE} \propto \sigma_{xx}^{1.6}$, near the quantum phase transition \cite{rf:30}.
Although WAL probes dephasing, while AHE reflects transverse transport. The coexistence of these anomalous scaling behaviors suggests that both may originate from a common quantum-critical modification of Dirac carriers.

We should mention that other scenarios involving a first-order transition, phase separation, and mass-gap formation have been proposed for $\alpha$-(BEDT-TTF)$_2$I$_3$ under pressure \cite{rf:31, rf:32, rf:33}. In such situations, a finite Dirac mass modifies the Berry phase and quantum interference corrections, potentially causing a crossover from WAL to weak localization (WL) in Dirac systems. This crossover has been theoretically discussed in topological surface states where a Dirac mass modifies the Berry phase and quantum interference correction \cite{rf:34}. Phase separation, on the other hand, would result in discontinuous changes in transport coefficients.

However, our magnetotransport results show that WAL persists continuously across $P_c$, with no sign of reversal or suppression of quantum interference. The $L_{\phi}$ remains large and evolves gradually with pressure, and the temperature exponent of dephasing exhibits a remarkable but continuous suppression. 
These observations strongly suggest a continuous quantum phase transition with gapless or nearly gapless Dirac excitations.

The long phase coherence length, $L_{\phi} \sim 700\text{–}800$ nm at 0.5 K, which remains nearly pressure-independent, indicates electronic homogeneity over at least submicron length scales. If substantial phase separation were present, domain boundaries would act as strong dephasing centers and significantly suppress $L_{\phi}$ to the domain scale. The absence of such a suppression rules out any macroscopic phase coexistence.

Finally, having established that the critical Dirac behavior and anomalous coherence scaling persist over a wide pressure range, we return to the high-pressure regime to examine the microscopic origin of the observed WAL, focusing on the role of spin–orbit interaction.

Within the HLN framework, the observation of WAL implies that spin–orbit scattering plays an important role in shaping the quantum interference correction. 
In particular, WAL emerges when spin–orbit scattering effectively suppresses triplet interference channels, leaving a dominant singlet contribution. 
This condition can be expressed as $\tau_{SO} \lesssim \tau_{\phi}$, corresponding to $L_{SO} \lesssim L_{\phi}$.

Using the experimentally obtained phase coherence length $L_{\phi} \sim 700$–800 nm at 0.5 K, we estimate the corresponding phase coherence time $\tau_{\phi}$ through the diffusion relation $L_{\phi} = \sqrt{D\tau_{\phi}}$. 
The diffusion constant is approximated as $D \approx v_F^2 \tau_{tr}/2$, appropriate for quasi-two-dimensional transport. 
Here, we adopt the avareaged in-plane Fermi velocity $v_F \sim 4.5 \times 10^4$ m/s \cite{rf:18, rf:35} and the transport scattering time $\tau_{tr} \sim 7 \times 10^{-11}$ s \cite{rf:36}.

These values yield $D \sim 0.07$–0.15 m$^2$/s and a phase coherence time $\tau_{\phi}$ on the order of a few picoseconds.
From the condition $\tau_{SO} \lesssim \tau_{\phi}$, we obtain a lower limit for the spin–orbit scattering energy scale of $E_{SO} \gtrsim \hbar/\tau_{\phi} \sim 0.1\text{–}0.3$ meV. 
This energy scale corresponds to a temperature of order 1–3 K, comparable to the experimental temperature range where WAL is observed. The SOC in the present system is not a negligible perturbation but instead provides an intrinsic low-energy scale that dominates quantum interference in this correlated Dirac system. 

In conclusion, we have investigated the weak antilocalization in the correlated Dirac semimetal $\alpha$-(BEDT-TTF)$_2$I$_3$ under pressure and revealed anomalous scaling of the phase coherence length across a quantum phase transition to a charge-ordered state. While the magnitude of $L_{\phi}$ remains remarkably robust ($700\text{–}800$ nm at 0.5 K), its temperature exponent is strongly suppressed near the critical pressure, indicating a breakdown of conventional diffusive Fermi-liquid dephasing. This anomalous scaling provides transport-based evidence for a quantum-critical Dirac regime subject to strong electron correlations.

The persistence of WAL across different phases strongly suggests that spin–orbit interaction plays a fundamental role in shaping quantum interference in this material. Together with the anomalous coherence scaling in the quantum-critical Dirac regime, this points to an intrinsic interplay between the critical Dirac physics and spin–orbit coupling.

These findings establish $\alpha$-(BEDT-TTF)$_2$I$_3$ as a unique platform for exploring quantum-critical transport in Dirac systems, where coherence, topology, and interaction effects are correlated.

\acknowledgments{
We thank T. Morinari at Kyoto University for valuable discussion and helpful comments. MEXT/JSPJ KAKENHI supported this work under Grant No. 24K06949.
}


\end{document}